\documentclass[sn-mathphys,Numbered]{sn-jnl}

\usepackage{graphicx}%
\usepackage{multirow}%
\usepackage{amsmath,amssymb,amsfonts}%
\usepackage{amsthm}%
\usepackage{mathrsfs}%
\usepackage[title]{appendix}%
\usepackage{xcolor}%
\usepackage{textcomp}%
\usepackage{manyfoot}%
\usepackage{booktabs}%
\usepackage{algorithm}%
\usepackage{algorithmicx}%
\usepackage{algpseudocode}%
\usepackage{listings}%

\usepackage{fancyvrb}

\DefineVerbatimEnvironment{Highlighting}{Verbatim}{commandchars=\\\{\}}
\newenvironment{Shaded}{\vspace{0.2cm}}{\vspace{0.2cm}}

\newcommand{\BuiltInTok}[1]{\textcolor[rgb]{0.00,0.50,0.00}{#1}}

\newcommand{\CommentTok}[1]{\textcolor[rgb]{0.38,0.63,0.69}{\textit{#1}}}

\newcommand{\ControlFlowTok}[1]{\textcolor[rgb]{0.00,0.44,0.13}{\textbf{#1}}}

\newcommand{\DecValTok}[1]{\textcolor[rgb]{0.25,0.63,0.44}{#1}}

\newcommand{\ExtensionTok}[1]{#1}

\newcommand{\FunctionTok}[1]{\textcolor[rgb]{0.02,0.16,0.49}{#1}}

\newcommand{\KeywordTok}[1]{\textcolor[rgb]{0.00,0.44,0.13}{\textbf{#1}}}
\newcommand{\NormalTok}[1]{#1}
\newcommand{\OperatorTok}[1]{\textcolor[rgb]{0.40,0.40,0.40}{#1}}

\newcommand{\StringTok}[1]{\textcolor[rgb]{0.25,0.44,0.63}{#1}}

\hypersetup{
  pdftitle={State of structural typing support in Scala 3.3.0},
  pdfauthor={Julien Richard-Foy},
  hidelinks}

\theoremstyle{thmstyleone}%
\theoremstyle{thmstyletwo}%

\theoremstyle{thmstylethree}%

\begin{document}

\title[State of structural typing support in Scala 3.3.0]{State of structural typing support in Scala 3.3.0}


\author{\fnm{Julien} \sur{Richard-Foy}}\email{julien@richard-foy.fr}


\abstract{Scala's type system is primarily based on nominal typing. Scala 3
introduces a special type, \texttt{Selectable}, which provides an
infrastructure for structural typing. Karlsson and Haller proposed
improvements to \texttt{Selectable} to support extensible records. In
this paper, we review several Scala 3 projects that involve structural
typing. We find that their implementation or usability would benefit
from the extensible records proposal from Karlsson and Haller. We
investigate the remaining common challenges when working with structural
types. In particular, we identify that a dedicated syntax for both
record types and record literals would be the most beneficial addition
to the language.}




\maketitle

\hypertarget{introduction}{%
\section{Introduction}\label{introduction}}

Nominal typing is very practical for most cases but, for some use cases,
such as data manipulation involving intermediate transformations, it is
not convenient. Indeed, defining a data type for every intermediate
result, even if it is used only once, is cumbersome. On the other hand,
structural typing is well suited for such cases as it allows developers
to manipulate the structure of data on-the-fly without having to declare
the type of every intermediate transformation.

A common workaround in Scala is to use a tuple. For instance, a method
performing a Euclidean division would typically be defined with a return
type of \texttt{(Int,\ Int)}, modeling a pair containing both the
quotient and the remainder. This solution has the benefit of being
concise. On the other hand, it is not obvious to the users of that
method which element of the pair is the quotient and which one is the
remainder. It is easy to confuse them. A solution to that drawback would
be to define a bespoke data type for the result of the Euclidean
division:

\begin{Shaded}
\begin{Highlighting}[]
\ControlFlowTok{case} \KeywordTok{class} \FunctionTok{EuclideanDivisionResult}\OperatorTok{(}\NormalTok{quotient}\OperatorTok{:} \BuiltInTok{Int}\OperatorTok{,}\NormalTok{ remainder}\OperatorTok{:} \BuiltInTok{Int}\OperatorTok{)}
\end{Highlighting}
\end{Shaded}

Thus, it would be obvious to the users which element is the quotient and
which one is the remainder. However, in practice, Scala developers do
not do follow this approach because it is too verbose compared to
returning a tuple.

Structural typing would bring the best of both worlds: implementers
would not need to define a data type such as
\texttt{EuclideanDivisionResult}, and yet users would get a value with
proper fields \texttt{quotient} and \texttt{remainder} preventing the
risk of confusion.

The goal of this paper is to identify the issues that prevent Scala
developers from using structural typing, and to propose solutions to
them.

In the next section, I describe the current support of structural typing
in Scala 3 based on the marker type \texttt{Selectable}, and I explain
why it does not provide a satisfactory solution to the above example.
Additionally, I present the language changes proposed by Karlsson and
Haller to support extensible records\cite{DBLP:conf/scala/KarlssonH18}, and I show how it improves the
situation.

In their paper, Karlsson and Haller evaluated the performance of their
records implementation. They showed that it is competitive with cached
reflection for structural field access. However, they did not validate
that their design provides a satisfactory solution to the use cases that
involve structural types. In section \ref{evaluation-of-extensible-records-on-real-world-libraries}, I review several existing
libraries that involve structural types (currently based on
\texttt{Selectable}), and I assess how much their implementation and
usage would be simplified by the extensible records proposal.

In section \ref{discussion}, I summarize the pros and cons of the extensible records
proposal, and I investigate the remaining common challenges regarding
both the implementation and usage of the reviewed libraries. My analysis
suggests that the most valuable addition to the extensible records
proposal would be a dedicated syntax for record types and record
literals.

The last section concludes.

In summary, the contributions of this work are the following:
\begin{itemize}
\item I evaluated the impact of the extensible records proposal on the
implementation and usage of several real world libraries,
\item I rebased the original implementation of the extensible records on the development
branch of the Scala 3.3.x series,
\item I proposed a new idea to improve further the support of structural typing in Scala 3: a dedicated syntax
for record types and record literals.
\end{itemize}

\hypertarget{support-of-structural-typing-in-scala-3}{%
\section{Support of Structural Typing in Scala
3}\label{support-of-structural-typing-in-scala-3}}

In this section, I describe the current support of structural typing in
Scala 3, and I present the extensible records proposal from Karlsson and
Haller.

\hypertarget{status-quo-the-marker-type-selectable}{%
\subsection{\texorpdfstring{Status Quo: the Marker Type
\texttt{Selectable}}{Status Quo: the Marker Type Selectable}}\label{status-quo-the-marker-type-selectable}}

Scala 3 comes with a special type, \texttt{Selectable}, which provides a
minimal infrastructure for structural typing.

As an example, the result of the Euclidean division can be modeled by
the type
\texttt{Record\ \{\ val\ quotient:\ Int;\ val\ remainder:\ Int\ \}},
provided that type \texttt{Record} is defined as a subtype of
\texttt{Selectable} that declares a method named \texttt{selectDynamic}:

\begin{Shaded}
\begin{Highlighting}[]
\KeywordTok{trait}\NormalTok{ Record }\KeywordTok{extends}\NormalTok{ Selectable}\OperatorTok{:}
  \KeywordTok{def} \FunctionTok{selectDynamic}\OperatorTok{(}\NormalTok{label}\OperatorTok{:} \ExtensionTok{String}\OperatorTok{):} \ExtensionTok{Any}
\end{Highlighting}
\end{Shaded}

Assuming we would have a value \texttt{result} of type
\texttt{Record\ \{\ val\ quotient:\ Int;\ val\ remainder:\ Int\ \}}, the
compiler would type the expression \texttt{result.quotient} as
\texttt{Int}, and the expression \texttt{result.remainder} as
\texttt{Int}. On the other hand, the compiler would reject any attempt
to select any other field, such as \texttt{result.fractional}, for
instance.

The way \texttt{Selectable} works is that the compiler only allows you
to select the fields that were declared in the type refinement (between
the braces following \texttt{Record}). It rewrites fields' selections to
calls to the method \texttt{selectDynamic}, which takes the name of the
selected field as a parameter of type \texttt{String}, and returns the
value of that field.

For end-users, \texttt{Selectable} only is half satisfactory. On the one
hand, once they get the result of the Euclidean division, they can
access the quotient and remainder in a concise and type-safe way. On the
other hand, the result type of the method is verbose, as illustrated by
the following snippet that compares a tuple type expression, a case
class reference, and a structural type expression:

\begin{Shaded}
\begin{Highlighting}[]
\OperatorTok{(}\BuiltInTok{Int}\OperatorTok{,} \BuiltInTok{Int}\OperatorTok{)}
\NormalTok{EuclideanDivisionResult}
\NormalTok{Record }\OperatorTok{\{} \KeywordTok{val}\NormalTok{ quotient}\OperatorTok{:} \BuiltInTok{Int}\OperatorTok{;} \KeywordTok{val}\NormalTok{ remainder}\OperatorTok{:} \BuiltInTok{Int} \OperatorTok{\}}
\end{Highlighting}
\end{Shaded}

As you can see, the structural type expression is the most verbose of
the three.

For the implementers of the method that performs the Euclidean division,
\texttt{Selectable} is not satisfactory at all because there is no
simple and type-safe way to create a structurally-typed value. The infrastructure in
the compiler is only responsible for rewriting fields' selections into
calls to \texttt{selectDynamic}, but that method still needs to be
implemented.

We could define \texttt{Record} as a class that internally stores
fields' values in a \texttt{Map} and implements \texttt{selectDynamic}
by performing a look-up in that \texttt{Map}:

\begin{Shaded}
\begin{Highlighting}[]
\ControlFlowTok{case} \KeywordTok{class} \FunctionTok{Record}\OperatorTok{(}\NormalTok{fields}\OperatorTok{:} \ExtensionTok{Map}\OperatorTok{[}\ExtensionTok{String}\OperatorTok{,} \ExtensionTok{Any}\OperatorTok{])} \KeywordTok{extends}\NormalTok{ Selectable}\OperatorTok{:}
  \KeywordTok{def} \FunctionTok{selectDynamic}\OperatorTok{(}\NormalTok{label}\OperatorTok{:} \ExtensionTok{String}\OperatorTok{):} \ExtensionTok{Any} \OperatorTok{=} \FunctionTok{fields}\OperatorTok{(}\NormalTok{label}\OperatorTok{)}
\end{Highlighting}
\end{Shaded}

With this \texttt{Record} data type defined, the implementers of the
method that performs the Euclidean division could return a
structurally-typed value as follows:

\begin{Shaded}
\begin{Highlighting}[]
\FunctionTok{Record}\OperatorTok{(}\ExtensionTok{Map}\OperatorTok{(}\StringTok{"quotient"} \OperatorTok{{-}\textgreater{}}\NormalTok{ quotient}\OperatorTok{,} \StringTok{"remainder"} \OperatorTok{{-}\textgreater{}}\NormalTok{ remainder}\OperatorTok{))}
  \OperatorTok{.}\NormalTok{asInstanceOf}\OperatorTok{[}\NormalTok{Record }\OperatorTok{\{} \KeywordTok{val}\NormalTok{ quotient}\OperatorTok{:} \BuiltInTok{Int}\OperatorTok{;} \KeywordTok{val}\NormalTok{ remainder}\OperatorTok{:} \BuiltInTok{Int} \OperatorTok{\}]}
\end{Highlighting}
\end{Shaded}

But that would still be much more verbose than returning a tuple or an
instance of the class \texttt{EuclideanDivisionResult}. Furthermore,
this is error-prone since any mistake in the type passed to the
\texttt{asInstanceOf} call would not be caught by the compiler (e.g., a
mismatch between the \texttt{Map} content and the declared record type).

Another important limitation in the current support of structural types
that is not illustrated by my example is that there is no way to
concatenate two structural types (for example to add new fields to a
record). The type intersection operation does almost that, but it can
not detect clashes (such as extending a record with a field of the same
name but with a different type). The limitations of structural types in
Scala 3 have been discussed in depth by Karlsson and Haller, and
motivated their proposal for extensible records, which I describe in the
next subsection.

\hypertarget{extensible-records-proposal}{%
\subsection{Extensible Records
Proposal}\label{extensible-records-proposal}}

As described in the previous section, \texttt{Selectable} alone does not
provide satisfactory support of structural typing in Scala 3. Karlsson
and Haller showed that the key missing piece is the ability to
\emph{extend} structurally-typed records in a type-safe way\cite{DBLP:conf/scala/KarlssonH18}. Indeed,
type-safe extension of records also provides type-safe \emph{creation}
of records since any record can be constructed by extending the empty
record with additional fields.

Here is how one can create a record containing a quotient and a
remainder with their proposal:

\begin{Shaded}
\begin{Highlighting}[]
\FunctionTok{Record}\OperatorTok{()} \OperatorTok{+} \OperatorTok{(}\StringTok{"quotient"} \OperatorTok{{-}\textgreater{}\textgreater{}}\NormalTok{ quotient}\OperatorTok{)} \OperatorTok{+} \OperatorTok{(}\StringTok{"remainder"} \OperatorTok{{-}\textgreater{}\textgreater{}}\NormalTok{ remainder}\OperatorTok{)}
\end{Highlighting}
\end{Shaded}

Or, with the specific constructor for records with two fields:

\begin{Shaded}
\begin{Highlighting}[]
\FunctionTok{Record}\OperatorTok{(}\StringTok{"quotient"} \OperatorTok{{-}\textgreater{}\textgreater{}}\NormalTok{ quotient}\OperatorTok{,} \StringTok{"remainder"} \OperatorTok{{-}\textgreater{}\textgreater{}}\NormalTok{ remainder}\OperatorTok{)}
\end{Highlighting}
\end{Shaded}

The type of both expressions is inferred to
\texttt{Record\ \{\ val\ quotient:\ Int;\ val\ remainder:\ Int\ \}}.

Without going into details about their proposal, the following points
are worth noting:

\begin{itemize}
\item to ensure soundness of record extension, a record
can only be extended by one field at a time (via the operator
\texttt{+}),

\item most of the proposal is implemented as a library but some
typing rules have been changed in the compiler to simplify the type
expression
\texttt{Record\ \{\ val\ a:\ A\ \}\ \&\ Record\ \{\ val\ b:\ B\ \}} into
\texttt{Record\ \{\ val\ a\ A;\ val\ b:\ B\ \}},

\item no special syntax is
introduced.
\end{itemize}

\hypertarget{evaluation-of-extensible-records-on-real-world-libraries}{%
\section{Evaluation of Extensible Records on Real World
Libraries}\label{evaluation-of-extensible-records-on-real-world-libraries}}

In this section, I look at three libraries that use structural typing,
and I evaluate how much the extensible records proposal would simplify
them.

I consider the following axes of simplification:
\begin{enumerate}
\item implementation: do
extensible records make those libraries easier to implement and
maintain? For instance, by subsuming some parts of their implementation.

\item usage: is the user-facing API easier to navigate through? Is it
easier to use? For instance, by having clearer type signatures, or a
more concise syntax.

\end{enumerate}

I selected the following libraries:

\begin{itemize}
\item Chimney, a library for
transforming data structures by adding, removing, or renaming fields,

\item Tyqu, a library for constructing SQL queries in a composable and
type-safe way,

\item Iskra, a library that adds more precise types to Spark
data frames.
\end{itemize}

I rebased the implementation of extensible records (published in 2018)
on top of Scala 3.3.0, re-implemented a small subset of Chimney, Tyqu,
and Iskra as self-contained tests, and looked at possible simplification
with records\footnote{The result is available at
\href{https://github.com/lampepfl/dotty/compare/main...julienrf:dotty:dotty-records-final}{https://github.com/lampepfl/dotty/compare/main\ldots julienrf:dotty:dotty-records-final}.}.

In the following subsections, I will give an overview of each library, I
will explain how they are implemented, and I will assess how much they
would be simplified by extensible records.

All the code examples assume following class definitions:

\begin{Shaded}
\begin{Highlighting}[]
\ControlFlowTok{case} \KeywordTok{class} \FunctionTok{UserV1}\OperatorTok{(}\NormalTok{name}\OperatorTok{:} \ExtensionTok{String}\OperatorTok{)}
\ControlFlowTok{case} \KeywordTok{class} \FunctionTok{UserV2}\OperatorTok{(}\NormalTok{name}\OperatorTok{:} \ExtensionTok{String}\OperatorTok{,}\NormalTok{ age}\OperatorTok{:} \ExtensionTok{Option}\OperatorTok{[}\BuiltInTok{Int}\OperatorTok{])}
\end{Highlighting}
\end{Shaded}

\hypertarget{chimney}{%
\subsection{Chimney}\label{chimney}}

Chimney transforms data structures by adding, removing, or renaming
fields. For instance, here is a snippet that converts a value of type
\texttt{UserV1} into a value of type \texttt{UserV2}:

\begin{Shaded}
\begin{Highlighting}[]
\KeywordTok{val}\NormalTok{ userV1 }\OperatorTok{=} \FunctionTok{UserV1}\OperatorTok{(}\StringTok{"Martin"}\OperatorTok{)}
\KeywordTok{val}\NormalTok{ userV2 }\OperatorTok{=}
\NormalTok{  userV1}
    \OperatorTok{.}\NormalTok{into}\OperatorTok{[}\NormalTok{UserV2}\OperatorTok{]}
    \OperatorTok{.}\FunctionTok{withFieldComputed}\OperatorTok{(}\NormalTok{\_}\OperatorTok{.}\NormalTok{age}\OperatorTok{,}\NormalTok{ \_ }\OperatorTok{=\textgreater{}} \BuiltInTok{None}\OperatorTok{)}
    \OperatorTok{.}\NormalTok{transform}
\end{Highlighting}
\end{Shaded}

We start with the value \texttt{userV1}, of type \texttt{UserV1}, and we
call the Chimney extension method \texttt{into} on it, with the type of
the target class \texttt{UserV2}.

Chimney tries to create a mapping from \texttt{UserV1} to
\texttt{UserV2} by looking at the fields in both classes that have the
same name and type. In this example, the field \texttt{name} of type
\texttt{String} is present in both classes, but the field \texttt{age}
is present only in the target class. Chimney creates a partial mapping
(handling only the field \texttt{name}), which we need to complete to
cover all the fields of the target class. We call
\texttt{withFieldComputed(\_.age,\ \_\ =\textgreater{}\ None)} to tell
Chimney to map the target field \texttt{age} to the value \texttt{None}.

Finally, when the mapping covers all the fields of the target class, we
can call the method \texttt{transform} to perform the conversion.

The current implementation of Chimney relies on macros that analyze the
structure of the source and target types to create the mapping between
them. The operation \texttt{withFieldComputed} is also implemented via a
\texttt{transparent\ inline\ def} that analyzes the first argument
(\texttt{\_.age} in the example above) to extract the name of the field.

On the usage side, the Chimney API is rather easy to navigate through
and IDEs provide helpful suggestions. For instance, it is impossible to
make typos on field names (e.g., \texttt{agee} instead of \texttt{age})
or assigning a value of an incompatible type to a field.

Furthermore, the authors of the library made a great job at providing
sensible error messages. For instance, if we omit to provide the mapping
for the \texttt{age} field:

\begin{Shaded}
\begin{Highlighting}[]
\NormalTok{Chimney can\textquotesingle{}t derive transformation from UserV1 to UserV2}

\NormalTok{UserV2}
\NormalTok{  age: Option[Int] {-} no accessor named age in source type UserV1}

\NormalTok{Consult https://chimney.readthedocs.io for usage examples.}
\end{Highlighting}
\end{Shaded}

There is a lot of overlap between what Chimney does and extensible
records. Indeed, with both tools developers transform data structures by
adding new fields to them.

Here is a possible sketch of how Chimney could look like if extensible
records were part of the language:

\begin{Shaded}
\begin{Highlighting}[]
\KeywordTok{val}\NormalTok{ userV2 }\OperatorTok{=}
  \OperatorTok{(}\NormalTok{userV1}\OperatorTok{.}\NormalTok{toRecord }\OperatorTok{+} \OperatorTok{(}\StringTok{"age"} \OperatorTok{{-}\textgreater{}\textgreater{}} \BuiltInTok{None}\OperatorTok{))}
    \OperatorTok{.}\NormalTok{into}\OperatorTok{[}\NormalTok{UserV2}\OperatorTok{]}
\end{Highlighting}
\end{Shaded}

We first convert the value \texttt{userV1} to an extensible record,
which we extend with a field \texttt{age}, and then we convert the
resulting record into the case class \texttt{UserV2}.

Thus, the implementation of Chimney would be only responsible for
providing conversions between case classes and records. The user-facing
API would be reduced to the following two extension methods:

\begin{Shaded}
\begin{Highlighting}[]
\NormalTok{extension }\OperatorTok{[}\NormalTok{A}\OperatorTok{](}\NormalTok{value}\OperatorTok{:}\NormalTok{ A}\OperatorTok{)}
  \KeywordTok{def}\NormalTok{ into}\OperatorTok{[}\NormalTok{B}\OperatorTok{](}\NormalTok{using Conversion}\OperatorTok{[}\NormalTok{A}\OperatorTok{,}\NormalTok{ B}\OperatorTok{]):}\NormalTok{ B }\OperatorTok{=}
\NormalTok{    value}\OperatorTok{.}\NormalTok{convert}
\NormalTok{extension }\OperatorTok{[}\NormalTok{A}\OperatorTok{](}\NormalTok{recordLike}\OperatorTok{:}\NormalTok{ A}\OperatorTok{)}
  \KeywordTok{def}\NormalTok{ toRecord}\OperatorTok{[}\NormalTok{B }\OperatorTok{\textless{}:}\NormalTok{ Record}\OperatorTok{](}\NormalTok{using Conversion}\OperatorTok{[}\NormalTok{A}\OperatorTok{,}\NormalTok{ B}\OperatorTok{]):}\NormalTok{ B }\OperatorTok{=}
\NormalTok{    recordLike}\OperatorTok{.}\NormalTok{convert}
\end{Highlighting}
\end{Shaded}

In addition to those methods, Chimney would still be responsible for
synthesizing \texttt{given} conversions between case classes and their
corresponding record type.

A drawback of the records-based approach is that typos and type-mismatch
errors in added fields would be caught only at the point of calling the
finalizer method \texttt{.into{[}UserV2{]}}.

\hypertarget{tyqu}{%
\subsection{Tyqu}\label{tyqu}}

Tyqu provides an embedded DSL to describe SQL queries from aggregation,
projection, and selection operations. Here is a snippet that shows how
to look up in a database the users whose name is ``Martin'':

\begin{Shaded}
\begin{Highlighting}[]
\KeywordTok{object}\NormalTok{ UsersV1 }\KeywordTok{extends}\NormalTok{ Table}\OperatorTok{:}
  \KeywordTok{val}\NormalTok{ name }\OperatorTok{=}\NormalTok{ Column}\OperatorTok{[}\ExtensionTok{String}\OperatorTok{]()}

\KeywordTok{val}\NormalTok{ results}\OperatorTok{:} \BuiltInTok{Seq}\OperatorTok{[}\ExtensionTok{Result} \OperatorTok{\{} \KeywordTok{def}\NormalTok{ firstName}\OperatorTok{:} \ExtensionTok{String} \OperatorTok{\}]} \OperatorTok{=}
  \FunctionTok{from}\OperatorTok{(}\NormalTok{UsersV1}\OperatorTok{)}
    \OperatorTok{.}\FunctionTok{filter}\OperatorTok{(}\NormalTok{user }\OperatorTok{=\textgreater{}}\NormalTok{ user}\OperatorTok{.}\NormalTok{name }\OperatorTok{===} \StringTok{"Martin"}\OperatorTok{)}
    \OperatorTok{.}\FunctionTok{map}\OperatorTok{(}\NormalTok{user }\OperatorTok{=\textgreater{}}\NormalTok{ user}\OperatorTok{.}\NormalTok{name}\OperatorTok{.}\FunctionTok{as}\OperatorTok{(}\StringTok{"firstName"}\OperatorTok{))}
    \OperatorTok{.}\FunctionTok{run}\OperatorTok{(}\NormalTok{dbConnection}\OperatorTok{)}
\end{Highlighting}
\end{Shaded}

We start by defining the schema of the \texttt{UsersV1} table via the
object of the same name. It has just one column, \texttt{name}, of type
\texttt{String}.

Then, we create a query that looks up in the \texttt{UsersV1} table by
calling \texttt{from(UsersV1)}, and we transform that query by calling
its \texttt{filter} and \texttt{map} methods. Ultimately, we execute the
query by calling the \texttt{run} method with an actual database
connection.

Tyqu provides a type-safe way to reference and manipulate table columns.
Both operations \texttt{filter} and \texttt{map} take as parameter a
function that takes a model of the table schema and returns a SQL
expression.

For example, we can refer to the column \texttt{name} by writing
\texttt{user.name}, which has type \texttt{Expression{[}String{]}}. This
lets us write the condition \texttt{user.name\ ===\ "Martin"}, which has
type \texttt{Expression{[}Boolean{]}} as expected by the \texttt{filter}
operation. Tyqu will translate those expressions into proper SQL when it
runs the query.

To achieve this, Tyqu synthesizes a type that models the table schema
(ie, the type of the \texttt{user} parameter). It does that via a macro
that introspects the content of the object \texttt{UsersV1} and finds
all its fields of type \texttt{Column}. It returns the structural type
\texttt{Selectable\ \{\ def\ name:\ Expression{[}String{]}\ \}}.

Another point worth noting is how projection operations can transform
the structural type of the underlying schema. For instance, in the call
to \texttt{map} we return \texttt{user.name.as("firstName")}, which
creates the alias \texttt{firstName} for the column \texttt{name}.
Subsequent transformations of the query would now have to use the
identifier \texttt{firstName} to refer to the column instead of
\texttt{name}.

When we run the query, Tyqu returns a result of the structural type
\texttt{Result\ \{\ def\ firstName:\ String\ \}} according to the alias
we created (the type \texttt{Result}, defined by Tyqu, is also a subtype
of \texttt{Selectable}).

There is some overlap between what Tyqu does and extensible records. The
most obvious one is the \texttt{Result} type defined by Tyqu, which is
literally the same as \texttt{Record}. Additionally, both Tyqu and
extensible records implement a way to create values of structural types.
Tyqu achieves that via the \texttt{as} operation (as in
\texttt{user.name.as("firstName")}).

Here is a possible sketch of how Tyqu could look like if extensible
records were part of the language:

\begin{Shaded}
\begin{Highlighting}[]
\KeywordTok{object}\NormalTok{ UsersV1Table}\OperatorTok{:}
  \KeywordTok{val}\NormalTok{ name }\OperatorTok{=}\NormalTok{ Column}\OperatorTok{[}\ExtensionTok{String}\OperatorTok{]()}

\KeywordTok{val}\NormalTok{ results}\OperatorTok{:} \BuiltInTok{Seq}\OperatorTok{[}\NormalTok{Record }\OperatorTok{\{} \KeywordTok{def}\NormalTok{ firstName}\OperatorTok{:} \ExtensionTok{String} \OperatorTok{\}]} \OperatorTok{=}
  \FunctionTok{from}\OperatorTok{(}\NormalTok{UsersV1Table}\OperatorTok{)}
    \OperatorTok{.}\FunctionTok{filter}\OperatorTok{(}\NormalTok{user }\OperatorTok{=\textgreater{}}\NormalTok{ user}\OperatorTok{.}\NormalTok{name }\OperatorTok{===} \StringTok{"Martin"}\OperatorTok{)}
    \OperatorTok{.}\FunctionTok{map}\OperatorTok{(}\NormalTok{user }\OperatorTok{=\textgreater{}} \FunctionTok{Record}\OperatorTok{(}\StringTok{"firstName"} \OperatorTok{{-}\textgreater{}\textgreater{}}\NormalTok{ user}\OperatorTok{.}\NormalTok{name}\OperatorTok{))}
    \OperatorTok{.}\FunctionTok{run}\OperatorTok{(}\NormalTok{connection}\OperatorTok{)}
\end{Highlighting}
\end{Shaded}

The differences with status quo are the following:

\begin{itemize}
\item executing a query
now returns standard records of type \texttt{Record} instead of
\texttt{Result},

\item the creation of the alias \texttt{firstName} is
achieved by constructing a standard record.
\end{itemize}

The benefits of the second points are debatable, as some users may
prefer using the original \texttt{as} method because it looks more
similar to SQL. I will elaborate further on that point in the Discussion
section.

Extensible records would not have a significant impact on the
implementation of Tyqu. Indeed, the main tasks related to structural
types in Tyqu consists of computing the type modeling table schemas from
table object definitions, and computing the type of a query result based
on the structural type of the query schema. Both tasks might be achieved
with some form of generic programming support on records, but such
features are not included in the extensible records proposal.

\hypertarget{iskra}{%
\subsection{Iskra}\label{iskra}}

Spark SQL allows developers to perform relational operations on data
structured in columns. A drawback of Spark SQL is that its main
abstraction, \texttt{DataFrame}, does not provide a precisely typed
schema of the manipulated data, making it easy to do mistakes such as
typos in column names, or applying an operation incompatible with the
underlying column type.

The goal of Iskra is to be as source compatible as possible with Spark's
\texttt{DataFrame} while providing more precise types in order to catch
the type errors mentioned above at compile-time. Here is a simple
example of Spark SQL program that prints the users whose name is
``Martin'', and then the same program with Iskra:

\begin{Shaded}
\begin{Highlighting}[]
\CommentTok{// Spark SQL}
\OperatorTok{(}\NormalTok{users}\OperatorTok{:}\NormalTok{ DataFrame}\OperatorTok{)}
  \OperatorTok{.}\FunctionTok{select}\OperatorTok{(}\NormalTok{$}\StringTok{"name"}\OperatorTok{.}\FunctionTok{as}\OperatorTok{(}\StringTok{"user\_name"}\OperatorTok{))}
  \OperatorTok{.}\FunctionTok{where}\OperatorTok{(}\NormalTok{$}\StringTok{"user\_name"} \OperatorTok{===} \StringTok{"Martin"}\OperatorTok{)}
  \OperatorTok{.}\FunctionTok{show}\OperatorTok{()}
\end{Highlighting}
\end{Shaded}

\begin{Shaded}
\begin{Highlighting}[]
\CommentTok{// Iskra}
\OperatorTok{(}\NormalTok{users}\OperatorTok{:}\NormalTok{ DataFrame}\OperatorTok{[}\NormalTok{UserV1}\OperatorTok{])}
  \OperatorTok{.}\FunctionTok{select}\OperatorTok{(}\NormalTok{$}\OperatorTok{.}\NormalTok{name}\OperatorTok{.}\FunctionTok{as}\OperatorTok{(}\StringTok{"user\_name"}\OperatorTok{))}
  \OperatorTok{.}\FunctionTok{where}\OperatorTok{(}\NormalTok{$}\OperatorTok{.}\NormalTok{user\_name }\OperatorTok{===} \StringTok{"Martin"}\OperatorTok{)}
  \OperatorTok{.}\FunctionTok{show}\OperatorTok{()}
\end{Highlighting}
\end{Shaded}

The only difference is that column selection is typechecked in Iskra:
the expression \texttt{\$.name} has type \texttt{Column{[}String{]}}.

Also shown in this example is the definition of column aliases, similar
to the Tyqu example. The result of the \texttt{select} operation is a
\texttt{DataFrame} with one column named \texttt{user\_name} and whose
type is \texttt{String}. As a consequence, we refer to that column in
the following \texttt{where} call by using \texttt{\$.user\_name}.

Iskra computes a structural type modeling the schema of the manipulated
\texttt{DataFrame} that is pretty similar to what Tyqu does. Iskra would
marginally benefit from extensible records, like Tyqu.

Here is a possible sketch of how Iskra could look like if extensible
records were part of the language:

\begin{Shaded}
\begin{Highlighting}[]
\OperatorTok{(}\NormalTok{users}\OperatorTok{:}\NormalTok{ DataFrame}\OperatorTok{[}\NormalTok{UserV1}\OperatorTok{])}
  \OperatorTok{.}\FunctionTok{select}\OperatorTok{(}\FunctionTok{Record}\OperatorTok{(}\StringTok{"user\_name"} \OperatorTok{{-}\textgreater{}\textgreater{}}\NormalTok{ $}\OperatorTok{.}\NormalTok{name}\OperatorTok{))}
  \OperatorTok{.}\FunctionTok{where}\OperatorTok{(}\NormalTok{$}\OperatorTok{.}\NormalTok{user\_name }\OperatorTok{===} \StringTok{"Martin"}\OperatorTok{)}
  \OperatorTok{.}\FunctionTok{show}\OperatorTok{()}
\end{Highlighting}
\end{Shaded}

Like with Tyqu, we use standard records to manipulate structurally-typed
values, which provides the same debatable benefit: the resulting
user-facing API would be pretty different from the original Spark SQL
API. I will discuss further that point in the next section.

The other main tasks related to structural types in Iskra consists of
computing the type modeling the \texttt{DataFrame} schema, and computing
the type of the result of evaluating a computation. Like with Tyqu, both
may be achievable with some support of generic programming on records,
but such a feature is not included in the extensible records proposal.

\hypertarget{discussion}{%
\section{Discussion}\label{discussion}}

In this section, I summarize the impacts of the extensible records
proposal on the three studied libraries, and I summarize the remaining
common challenges when working with structural types. Then, I make new
proposals to address those challenges.

\hypertarget{pros-and-cons-of-extensible-records}{%
\subsection{Pros and Cons of Extensible
Records}\label{pros-and-cons-of-extensible-records}}

The following table summarizes the impact of extensible records on the
studied libraries (``+'' means a marginal positive impact, and ``++''
means a significant positive impact):

\begin{table}[h]
\begin{tabular}{@{}llll@{}}
\toprule
& Chimney & Tyqu & Iskra \\
\midrule
Implementation & ++ & + & + \\
Usage & = & + & + \\
\botrule
\end{tabular}
\end{table}

\hypertarget{impact-on-implementation}{%
\subsubsection{Impact on
Implementation}\label{impact-on-implementation}}

The only case where the benefits are clear is the implementation of
Chimney. Indeed, 80\% of Chimney's implementation is a DSL to extend
data structures. That part is subsumed by the extensible records
proposal. The remaining 20\% is an infrastructure to convert between
case classes and records.

The impact of the extensible records on the implementation of Tyqu and
Iskra is only marginal. Indeed, most of the work related to structural
types performed by Tyqu and Iskra consists of computing the types of SQL
table schemas (or \texttt{DataFrame} schemas, respectively), and
computing the result types of running the queries (or evaluating the
computations, respectively).

For example, in Iskra, calling the \texttt{map} method with a function
that returns a record of type
\texttt{Record\ \{\ val\ a:\ Column{[}A{]};\ val\ b:\ Column{[}B{]}\ \}}
produces a
\texttt{DataFrame{[}Schema\ \{\ val\ a:\ Column{[}A{]};\ val\ b:\ Column{[}B{]}\ \}{]}}.
ie, the schema of the resulting \texttt{DataFrame} has the same columns
as the provided record. Note that it might be possible to just reuse the
\texttt{Record} type itself to model a schema instead of using a bespoke
\texttt{Schema} type, but the current implementation in Iskra requires
the \texttt{Schema} type to have specific additional methods that are
not on \texttt{Record}.

Likewise, evaluating a computation on a
\texttt{DataFrame{[}Schema\ \{\ val\ a:\ Column{[}A{]};\ val\ b:\ Column{[}B{]}\ \}{]}}
produces a result of type
\texttt{Seq{[}Record\ \{\ val\ a:\ A;\ val\ b:\ B\ \}{]}}. Here, the
structural type of the result has the same field names as the
\texttt{DataFrame} schema, but their types do not contain the
\texttt{Column} type constructor anymore.

Similar examples could be constructed for Tyqu.

Unfortunately, the extensible records proposal is of no help to compute
such types.

\hypertarget{impact-on-usage}{%
\subsubsection{Impact on Usage}\label{impact-on-usage}}

In all the cases, the impact of extensible records on the usability of
the libraries is marginal.

Nevertheless, a general improvement is that once records become standard
in Scala, then there is a unique, standard way to create and extend
records that is shared by all the libraries. The user experience is more
consistent across the libraries, and fewer concepts need to be learned
to get started with a new library.

Other than that, the current design of the extensible records proposal
still suffers from usability issues.

A minor issue is that in the proposed design, records can only be
extended by using the \texttt{+} symbolic operation. This does not work
well when mixed with non-infix calls. Compare for instance the following
typical usage of Chimney:

\begin{Shaded}
\begin{Highlighting}[]
\NormalTok{foo}
  \OperatorTok{.}\NormalTok{into}\OperatorTok{[}\NormalTok{Bar}\OperatorTok{]}
  \OperatorTok{.}\FunctionTok{withFieldConst}\OperatorTok{(}\NormalTok{\_}\OperatorTok{.}\NormalTok{baz}\OperatorTok{,} \DecValTok{42}\OperatorTok{)}
  \OperatorTok{.}\FunctionTok{withFieldComputed}\OperatorTok{(}\NormalTok{\_}\OperatorTok{.}\NormalTok{quux}\OperatorTok{,}\NormalTok{ \_}\OperatorTok{.}\NormalTok{x }\OperatorTok{+} \DecValTok{1}\OperatorTok{)}
  \OperatorTok{.}\NormalTok{transform}
\end{Highlighting}
\end{Shaded}

With the following equivalent program based on my records-based
adaptation of Chimney:

\begin{Shaded}
\begin{Highlighting}[]
\OperatorTok{(}\NormalTok{foo}
  \OperatorTok{.}\NormalTok{toRecord}
    \OperatorTok{+} \OperatorTok{(}\StringTok{"baz"} \OperatorTok{{-}\textgreater{}\textgreater{}} \DecValTok{42}\OperatorTok{)}
    \OperatorTok{+} \OperatorTok{(}\StringTok{"quux"} \OperatorTok{{-}\textgreater{}\textgreater{}}\NormalTok{ foo}\OperatorTok{.}\NormalTok{x }\OperatorTok{+} \DecValTok{1}\OperatorTok{))}
  \OperatorTok{.}\NormalTok{into}\OperatorTok{[}\NormalTok{Bar}\OperatorTok{]}
\end{Highlighting}
\end{Shaded}

The combination of infix operators and dot-notation does not play well.

This problem can easily be solved by providing a non-symbolic alias to
the operation \texttt{+}, as I will show in section \ref{non-symbolic-record-extension-operation}.

Another, more important issue with the extensible records design is that
the syntax to define a field is not intuitive. To define a field name,
developers write a \texttt{String} literal. This is inconsistent with
the usual way of creating bindings in Scala, which consists of writing
an unquoted identifier in a binding position. Fixing this problem would
require changes to the language. Obviously, the current implementations
of Tyqu and Iskra also suffer from this problem.

The last important issue is also related to syntax. As shown in the
introduction, the syntax of structural types is very verbose. As a
reminder, here is the type of records containing the quotient and
remainder of a division:

\begin{Shaded}
\begin{Highlighting}[]
\NormalTok{Record }\OperatorTok{\{} \KeywordTok{val}\NormalTok{ quotient}\OperatorTok{:} \BuiltInTok{Int}\OperatorTok{;} \KeywordTok{val}\NormalTok{ remainder}\OperatorTok{:} \BuiltInTok{Int} \OperatorTok{\}}
\end{Highlighting}
\end{Shaded}

It is obviously much more verbose than a tuple containing the same data.

This problem is important since, unlike classes, structural types can
not be referred to by a simple identifier such as
\texttt{EuclideanDivisionResult}. Defining a type alias would not really
solve this issue because structural type expressions would still show up
in signatures in the API documentation or in IDEs. Structural types are
meant to be referred to by their type expression. Defining such an alias
would defeat their purpose.

In the next section, I propose some ideas to fix this issue and the
other aforementioned issues.

Finally, another minor usability issue with the current implementation
of extensible records is that the type name \texttt{Record} clashes with
\texttt{java.lang.Record}, introduced in JDK 16.

\hypertarget{proposals-to-address-the-remaining-issues}{%
\subsection{Proposals to Address the Remaining
Issues}\label{proposals-to-address-the-remaining-issues}}

I propose three possible additions to the original extensible records
proposal to address the issues described in the previous section: a
non-symbolic operation to extend records, a dedicated syntax for record
types and record literals, and compiler-synthesized mirrors for record
types.

\hypertarget{non-symbolic-record-extension-operation}{%
\subsubsection{Non-Symbolic Record Extension
Operation}\label{non-symbolic-record-extension-operation}}

As mentioned in the previous section, in the extensible records
proposal, the only way to extend a record is to use a symbolic
\texttt{+} operation, which does not play well with the dot-notation. We
can fix this issue by adding a non-symbolic alias \texttt{withField} to
the \texttt{+} operation on records. With this change, the initial example
becomes:

\begin{Shaded}
\begin{Highlighting}[]
\NormalTok{foo}
  \OperatorTok{.}\NormalTok{toRecord}
  \OperatorTok{.}\FunctionTok{withField}\OperatorTok{(}\StringTok{"baz"} \OperatorTok{{-}\textgreater{}\textgreater{}} \DecValTok{42}\OperatorTok{)}
  \OperatorTok{.}\FunctionTok{withField}\OperatorTok{(}\StringTok{"quux"} \OperatorTok{{-}\textgreater{}\textgreater{}}\NormalTok{ foo}\OperatorTok{.}\NormalTok{x }\OperatorTok{+} \DecValTok{1}\OperatorTok{)}
  \OperatorTok{.}\NormalTok{into}\OperatorTok{[}\NormalTok{Bar}\OperatorTok{]}
\end{Highlighting}
\end{Shaded}

\hypertarget{dedicated-syntax}{%
\subsubsection{Dedicated Syntax}\label{dedicated-syntax}}

The syntax for records should be concise, readable, and lead to as few
ambiguities as possible.

The syntax for record types and record literals should be homogeneous,
just like the syntax of tuple types and tuple literals, and the syntax
of case class definitions and case class construction.

Tuples, records, and case classes are all means of structuring
information. Ideally, the syntax of records should share similarities
with the syntax of tuples and case classes.

With these principles in mind, I propose the following syntax to define
a record containing the quotient and remainder of a Euclidean division:

\begin{Shaded}
\begin{Highlighting}[]
\OperatorTok{(}\NormalTok{quotient}\OperatorTok{:} \BuiltInTok{Int}\OperatorTok{,}\NormalTok{ remainder}\OperatorTok{:} \BuiltInTok{Int}\OperatorTok{)}
\end{Highlighting}
\end{Shaded}

This type would expand to the following:

\begin{Shaded}
\begin{Highlighting}[]
\NormalTok{Record }\OperatorTok{\{} \KeywordTok{val}\NormalTok{ quotient}\OperatorTok{:} \BuiltInTok{Int}\OperatorTok{;} \KeywordTok{val}\NormalTok{ remainder}\OperatorTok{:} \BuiltInTok{Int} \OperatorTok{\}}
\end{Highlighting}
\end{Shaded}

Constructing a record literal of that type looks as follows:

\begin{Shaded}
\begin{Highlighting}[]
\OperatorTok{(}\NormalTok{quotient }\OperatorTok{=}\NormalTok{ dividend }\OperatorTok{/}\NormalTok{ divisor}\OperatorTok{,}\NormalTok{ remainder }\OperatorTok{=}\NormalTok{ dividend }\OperatorTok{\%}\NormalTok{ divisor}\OperatorTok{)}
\end{Highlighting}
\end{Shaded}

Extending a record looks as follows:

\begin{Shaded}
\begin{Highlighting}[]
\OperatorTok{(}\NormalTok{foo }\OperatorTok{=} \DecValTok{42}\OperatorTok{)} \OperatorTok{+} \OperatorTok{(}\NormalTok{bar }\OperatorTok{=} \StringTok{"hello"}\OperatorTok{)}
\end{Highlighting}
\end{Shaded}

Or, with the non-symbolic operator:

\begin{Shaded}
\begin{Highlighting}[]
\OperatorTok{(}\NormalTok{foo }\OperatorTok{=} \DecValTok{42}\OperatorTok{)}\OperatorTok{.}\FunctionTok{withField}\OperatorTok{((}\NormalTok{bar }\OperatorTok{=} \StringTok{"hello"}\OperatorTok{))}
\end{Highlighting}
\end{Shaded}

Just like auto-tupling allows developers to call a function taking a
tuple as if it was a function taking several parameters,
``auto-recording'' allows developers to call a function taking a record
as if it was a function taking several named parameters:

\begin{Shaded}
\begin{Highlighting}[]
\OperatorTok{(}\NormalTok{foo }\OperatorTok{=} \DecValTok{42}\OperatorTok{)}\OperatorTok{.}\FunctionTok{withField}\OperatorTok{(}\NormalTok{bar }\OperatorTok{=} \StringTok{"hello"}\OperatorTok{)}
\end{Highlighting}
\end{Shaded}

``auto-recording'' can be a source of ambiguities, though, and more work
is needed to find the right set of rules governing its application.

Another benefit of having a dedicated syntax for record literals is that
it allows the extension of a record with more than one field at a time,
as in:

\begin{Shaded}
\begin{Highlighting}[]
\OperatorTok{(}\NormalTok{foo }\OperatorTok{=} \DecValTok{42}\OperatorTok{)} \OperatorTok{+} \OperatorTok{(}\NormalTok{bar }\OperatorTok{=} \StringTok{"hello"}\OperatorTok{,}\NormalTok{ baz }\OperatorTok{=} \KeywordTok{true}\OperatorTok{)}
\end{Highlighting}
\end{Shaded}

This was not possible in the original extensible records proposal
because of soundness issues.

Note that extending a record with a non-literal would still be
forbidden:

\begin{Shaded}
\begin{Highlighting}[]
\KeywordTok{val}\NormalTok{ r }\OperatorTok{=} \OperatorTok{(}\NormalTok{bar }\OperatorTok{=} \StringTok{"hello"}\OperatorTok{,}\NormalTok{ baz }\OperatorTok{=} \KeywordTok{true}\OperatorTok{)}
\OperatorTok{(}\NormalTok{foo }\OperatorTok{=} \DecValTok{42}\OperatorTok{)} \OperatorTok{+}\NormalTok{ r }\CommentTok{// error}
\end{Highlighting}
\end{Shaded}

Last, records should also support named field patterns, as in
\href{https://github.com/scala/improvement-proposals/pull/44}{SIP-44}.

With this syntax, the code examples presented earlier can be rewritten
as follows:

\begin{Shaded}
\begin{Highlighting}[]
\NormalTok{userV1}
  \OperatorTok{.}\NormalTok{toRecord}
  \OperatorTok{.}\FunctionTok{withField}\OperatorTok{(}\NormalTok{age }\OperatorTok{=} \BuiltInTok{None}\OperatorTok{)}
  \OperatorTok{.}\NormalTok{into}\OperatorTok{[}\NormalTok{UserV2}\OperatorTok{]}
\end{Highlighting}
\end{Shaded}

\begin{Shaded}
\begin{Highlighting}[]
\NormalTok{foo}
  \OperatorTok{.}\NormalTok{toRecord}
  \OperatorTok{.}\FunctionTok{withField}\OperatorTok{(}\NormalTok{baz }\OperatorTok{=} \DecValTok{42}\OperatorTok{)}
  \OperatorTok{.}\FunctionTok{withField}\OperatorTok{(}\NormalTok{quux }\OperatorTok{=}\NormalTok{ foo}\OperatorTok{.}\NormalTok{x }\OperatorTok{+} \DecValTok{1}\OperatorTok{)}
  \OperatorTok{.}\NormalTok{into}\OperatorTok{[}\NormalTok{Bar}\OperatorTok{]}
\end{Highlighting}
\end{Shaded}

\begin{Shaded}
\begin{Highlighting}[]
\KeywordTok{val}\NormalTok{ results}\OperatorTok{:} \BuiltInTok{Seq}\OperatorTok{[(}\NormalTok{firstName}\OperatorTok{:} \ExtensionTok{String}\OperatorTok{)]} \OperatorTok{=}
  \FunctionTok{from}\OperatorTok{(}\NormalTok{UsersV1Table}\OperatorTok{)}
    \OperatorTok{.}\FunctionTok{filter}\OperatorTok{(}\NormalTok{user }\OperatorTok{=\textgreater{}}\NormalTok{ user}\OperatorTok{.}\NormalTok{name }\OperatorTok{===} \StringTok{"Martin"}\OperatorTok{)}
    \OperatorTok{.}\FunctionTok{map}\OperatorTok{(}\NormalTok{user }\OperatorTok{=\textgreater{}} \OperatorTok{(}\NormalTok{firstName }\OperatorTok{=}\NormalTok{ user}\OperatorTok{.}\NormalTok{name}\OperatorTok{))}
    \OperatorTok{.}\FunctionTok{run}\OperatorTok{(}\NormalTok{connection}\OperatorTok{)}
\end{Highlighting}
\end{Shaded}

\begin{Shaded}
\begin{Highlighting}[]
\NormalTok{users}
  \OperatorTok{.}\FunctionTok{select}\OperatorTok{(}\NormalTok{user\_name }\OperatorTok{=}\NormalTok{ $}\OperatorTok{.}\NormalTok{name}\OperatorTok{)}
  \OperatorTok{.}\FunctionTok{show}\OperatorTok{()}
\end{Highlighting}
\end{Shaded}

I believe the simpler type signatures and literal syntax would make it
easier to work with structural types.

An open question is whether the dedicated syntax should apply to records
only, or to structural types in general. Even if records become standard
in Scala, the underlying, more general, mechanism based on
\texttt{Selectable} supports other forms of structurally-typed values.
Probably, a more general syntax that would also apply to any form of
structural type would be useful. A possibility could be to support a
shorthand syntax for type refinements: the type
\texttt{A(x:\ X,\ y:\ Y)} would be a shorthand for
\texttt{A\ \{\ val\ x:\ X;\ val\ y:\ Y\ \}}.

\hypertarget{generic-programming-support-for-records}{%
\subsubsection{Generic Programming Support for
Records}\label{generic-programming-support-for-records}}

The last challenge related to using structural types, faced by both Tyqu
and Iskra, is to derive a structural type from another (possibly
structural) type.

Let us remind one of the examples presented earlier. In Iskra, the type
\texttt{DataFrame{[}Schema\ \{\ val\ a:\ Column{[}A{]};\ val\ b:\ Column{[}B{]}\ \}{]}}
describes a computation that produces results of type
\texttt{Record\ \{\ val\ a:\ A;\ val\ b:\ B\ \}}. Computing the result
type is mechanical: for every field \texttt{f:\ Column{[}T{]}} of the
schema type, there is a corresponding field \texttt{f:\ T} in the
resulting record. A generic solution to perform this type computation
requires a way to iterate over the fields of a structural type.

In Scala 3, mirrors allow developers to iterate over the fields of case
classes. However, this technique can not currently be used with
structural types because mirrors are not synthesized for them. In the
remainder of this section, I propose new rules to synthesize
mirrors for structural types and assess how much that would simplify the
implementation of Iskra, Tyqu, and Chimney.

Currently, the compiler synthesizes a subtype of \texttt{Mirror.Product}
for every case class definition:

\begin{Shaded}
\begin{Highlighting}[]
\KeywordTok{trait}\NormalTok{ Product }\KeywordTok{extends}\NormalTok{ Mirror}\OperatorTok{:}
  \KeywordTok{type}\NormalTok{ MirroredElemLabels }\OperatorTok{\textless{}:}\NormalTok{ Tuple}
  \KeywordTok{type}\NormalTok{ MirroredElemTypes }\OperatorTok{\textless{}:}\NormalTok{ Tuple}
  \KeywordTok{type}\NormalTok{ MirroredLabel }\OperatorTok{\textless{}:} \ExtensionTok{String}
  \KeywordTok{type}\NormalTok{ MirroredMonoType}
  \KeywordTok{def} \FunctionTok{fromProduct}\OperatorTok{(}\NormalTok{product}\OperatorTok{:}\NormalTok{ scala}\OperatorTok{.}\NormalTok{Product}\OperatorTok{):}\NormalTok{ MirroredMonoType}
\end{Highlighting}
\end{Shaded}

For instance, here is a case class definition and its mirror type as
synthesized by the compiler:

\begin{Shaded}
\begin{Highlighting}[]
\ControlFlowTok{case} \KeywordTok{class} \FunctionTok{EuclideanDivisionResult}\OperatorTok{(}\NormalTok{quotient}\OperatorTok{:} \BuiltInTok{Int}\OperatorTok{,}\NormalTok{ remainder}\OperatorTok{:} \BuiltInTok{Int}\OperatorTok{)}

\CommentTok{// Synthesized by the compiler}
\KeywordTok{trait}\NormalTok{ EuclideanDivisionResultMirror }\KeywordTok{extends}\NormalTok{ Mirror}\OperatorTok{.}\NormalTok{Product}\OperatorTok{:}
  \KeywordTok{type}\NormalTok{ MirroredElemLabels }\OperatorTok{=} \OperatorTok{(}\StringTok{"quotient"}\OperatorTok{,} \StringTok{"remainder"}\OperatorTok{)}
  \KeywordTok{type}\NormalTok{ MirroredElemTypes  }\OperatorTok{=} \OperatorTok{(}\BuiltInTok{Int}\OperatorTok{,} \BuiltInTok{Int}\OperatorTok{)}
  \KeywordTok{type}\NormalTok{ MirroredLabel      }\OperatorTok{=} \StringTok{"EuclideanDivisionResult"}
  \KeywordTok{type}\NormalTok{ MirroredMonoType   }\OperatorTok{=}\NormalTok{ EuclideanDivisionResult}
  \KeywordTok{def} \FunctionTok{fromProduct}\OperatorTok{(}\NormalTok{product}\OperatorTok{:}\NormalTok{ scala}\OperatorTok{.}\NormalTok{Product}\OperatorTok{):}\NormalTok{ EuclideanDivisionResult }\OperatorTok{=}
    \FunctionTok{EuclideanDivisionResult}\OperatorTok{(}
\NormalTok{      product}\OperatorTok{.}\FunctionTok{productElement}\OperatorTok{(}\DecValTok{0}\OperatorTok{).}\NormalTok{asInstanceOf}\OperatorTok{[}\BuiltInTok{Int}\OperatorTok{],}
\NormalTok{      product}\OperatorTok{.}\FunctionTok{productElement}\OperatorTok{(}\DecValTok{1}\OperatorTok{).}\NormalTok{asInstanceOf}\OperatorTok{[}\BuiltInTok{Int}\OperatorTok{]}
    \OperatorTok{)}
\end{Highlighting}
\end{Shaded}

The type members of \texttt{EuclideanDivisionResultMirror} model the
structure of the class \texttt{EuclideanDivisionResult}. The mirror also
implements the method \texttt{fromProduct}, which provides a constructor
for \texttt{EuclideanDivisionResult}. It is worth noting that this
constructor is position-based: it iterates over the elements of the
provided product and passes them as parameters to the primary
constructor of the class \texttt{EuclideanDivisionResult}.

That mechanics would not work as is on structural types because position
is not significant in structural types. Instead, the constructor should
be name-based:

\begin{Shaded}
\begin{Highlighting}[]
\KeywordTok{def} \FunctionTok{fromFields}\OperatorTok{(}\NormalTok{fields}\OperatorTok{:} \ExtensionTok{Map}\OperatorTok{[}\ExtensionTok{String}\OperatorTok{,} \ExtensionTok{Any}\OperatorTok{]):}\NormalTok{ MirroredMonoType}
\end{Highlighting}
\end{Shaded}

The complete template type for record type mirrors would be the
following:

\begin{Shaded}
\begin{Highlighting}[]
\KeywordTok{object}\NormalTok{ Mirror}\OperatorTok{:}
  \KeywordTok{trait}\NormalTok{ Record }\KeywordTok{extends}\NormalTok{ Mirror}\OperatorTok{:}
    \KeywordTok{type}\NormalTok{ MirroredElemLabels }\OperatorTok{\textless{}:}\NormalTok{ Tuple}
    \KeywordTok{type}\NormalTok{ MirroredElemTypes }\OperatorTok{\textless{}:}\NormalTok{ Tuple}
    \KeywordTok{type}\NormalTok{ MirroredLabel }\OperatorTok{=}\NormalTok{ Nothing}
    \KeywordTok{type}\NormalTok{ MirroredMonoType }\OperatorTok{\textless{}:}\NormalTok{ scala}\OperatorTok{.}\NormalTok{Record}
    \KeywordTok{def} \FunctionTok{fromFields}\OperatorTok{(}\NormalTok{fields}\OperatorTok{:} \ExtensionTok{Map}\OperatorTok{[}\ExtensionTok{String}\OperatorTok{,} \ExtensionTok{Any}\OperatorTok{]:}\NormalTok{ MirroredMonoType}
\end{Highlighting}
\end{Shaded}

Then, the compiler would synthesize the following mirror type for the
record type modeling the result of a Euclidean division:

\begin{Shaded}
\begin{Highlighting}[]
\KeywordTok{trait}\NormalTok{ RecordQuotientRemainderMirror }\KeywordTok{extends}\NormalTok{ Mirror}\OperatorTok{.}\NormalTok{Record}\OperatorTok{:}
  \KeywordTok{type}\NormalTok{ MirroredElemLabels }\OperatorTok{=} \OperatorTok{(}\StringTok{"quotient"}\OperatorTok{,} \StringTok{"remainder"}\OperatorTok{)}
  \KeywordTok{type}\NormalTok{ MirroredElemTypes }\OperatorTok{=} \OperatorTok{(}\BuiltInTok{Int}\OperatorTok{,} \BuiltInTok{Int}\OperatorTok{)}
  \KeywordTok{type}\NormalTok{ MirroredMonoType }\OperatorTok{=}
    \NormalTok{ Record }\OperatorTok{\{} \KeywordTok{val}\NormalTok{ quotient}\OperatorTok{:} \BuiltInTok{Int}\OperatorTok{;} \KeywordTok{val}\NormalTok{ remainder}\OperatorTok{:} \BuiltInTok{Int} \OperatorTok{\}}
  \KeywordTok{def} \FunctionTok{fromFields}\OperatorTok{(}
    \NormalTok{fields}\OperatorTok{:} \ExtensionTok{Map}\OperatorTok{[}\ExtensionTok{String}\OperatorTok{,} \ExtensionTok{Any}\OperatorTok{]}
  \OperatorTok{):}\NormalTok{ Record }\OperatorTok{\{} \KeywordTok{val}\NormalTok{ quotient}\OperatorTok{:} \BuiltInTok{Int}\OperatorTok{;} \KeywordTok{val}\NormalTok{ remainder}\OperatorTok{:} \BuiltInTok{Int} \OperatorTok{\}} \OperatorTok{=}
    \FunctionTok{Record}\OperatorTok{()}
      \OperatorTok{+} \OperatorTok{(}\StringTok{"quotient"} \OperatorTok{{-}\textgreater{}\textgreater{}} \FunctionTok{fields}\OperatorTok{(}\StringTok{"quotient"}\OperatorTok{).}\NormalTok{asInstanceOf}\OperatorTok{[}\BuiltInTok{Int}\OperatorTok{])}
      \OperatorTok{+} \OperatorTok{(}\StringTok{"remainder"} \OperatorTok{{-}\textgreater{}\textgreater{}} \FunctionTok{fields}\OperatorTok{(}\StringTok{"remainder"}\OperatorTok{).}\NormalTok{asInstanceOf}\OperatorTok{[}\BuiltInTok{Int}\OperatorTok{])}
\end{Highlighting}
\end{Shaded}

Note that such mirrors could not be synthesized for \texttt{Selectable}
in general, because \texttt{Selectable} also supports structural method
calls via a method \texttt{applyDynamic}, not just structural field
selection via \texttt{selectDynamic}.

The benefits of record mirrors would be to provide a standard,
high-level, way to iterate over the fields of a record type.

Currently, it is already possible to iterate over the fields of
structural types by iterating over the \texttt{Refinement} nodes of the
\texttt{TypeRepr} model of the structural type, with TASTy reflection.
However, such refinements may also include arbitrary refinements such as
\texttt{\{\ type\ X;\ def\ f(x:\ Int):\ X\ \}}.

Record mirrors would simplify the work of the developers by removing the
need to check the structure of the refinements: the compiler would
synthesize a mirror only if the refinements all have the form
\texttt{val\ \textless{}name\textgreater{}:\ \textless{}type\textgreater{}},
and that none of the field types uses another field name as a type
prefix (as in \texttt{\{\ val\ x:\ X;\ val\ y:\ x.Y\ \}}).

Even though our motivating example only mentions Iskra and Tyqu, it is
likely that record mirrors would also benefit to a library like Chimney
to implement the conversions from records to case classes. The
developers would summon both the mirror of the source record type and
the target case class, check that all the fields of the case class are
present in the record type and have a compatible type, and finally
construct the case class instance by retrieving all the fields' values
by calling \texttt{selectDynamic} on the underlying record instance.

That being said, record mirrors would also have some drawbacks.
Unlike case class mirrors, which have almost no compile-time and
run-time footprint, record mirror types cannot be cached by the compiler
as efficiently as case class mirrors, and their usage would require the
allocation of extra objects.

In conclusion, record mirrors may provide additional benefits on the
implementation of libraries manipulating structural types, but more work
is necessary to more precisely assess their impact.

\hypertarget{conclusion}{%
\section{Conclusion}\label{conclusion}}

In this paper, I reviewed the support of structural typing in Scala
3.3.0 and identified possible ways to improve it.

The first issue is that there is no standard and type-safe way to create
structurally-typed values. I assessed that that issue is fully solved by
the extensible records proposal from Karlsson and Haller.

My analysis showed that the next most important issue is that the syntax
of structural type expressions is too verbose. Structural types, by
definition, do not have a short name that can be used to reference them.
Therefore, they need a concise syntax to be usable at scale. I proposed
a syntax for record type expressions, and a consistent syntax for record
literals.

I also investigated other minor issues and their possible solutions,
such as generic programming on records via compiler-synthesized mirrors.


\bibliography{main}

\end{document}